\documentclass[12pt]{article}

\usepackage[T1]{fontenc}
\usepackage[utf8]{inputenc}
\usepackage{lmodern}
\usepackage{microtype}
\usepackage{amsmath,amssymb,bm}
\usepackage{graphicx}
\usepackage{booktabs}
\usepackage{tabularx}
\usepackage{longtable}
\usepackage{enumitem}
\usepackage{placeins}
\usepackage{float}
\usepackage[margin=1in]{geometry}
\usepackage[numbers,sort&compress]{natbib}
\usepackage[hidelinks]{hyperref}
\usepackage[nameinlink,noabbrev]{cleveref}

\graphicspath{{figures/}}

\setlist{nosep}
\providecommand{\tightlist}{\setlength{\itemsep}{0pt}\setlength{\parskip}{0pt}}

\newcommand{\dd}{\mathrm{d}}

\title{Direct Optimization of Stellarator Omnigenity from the Second Adiabatic Invariant}
\author{Hanlin Chen$^{1,2}$, Zhiyuan Lu$^{1,*}$, Guosheng Xu$^{1,*}$, Shuai Cao$^{1,2}$,\\
  Yang Han$^{1,2}$, Dehong Chen$^1$, and Baonian Wan$^1$\\[0.4em]
  \small $^1$Institute of Plasma Physics, Chinese Academy of Sciences, Hefei 230031, China\\
  \small $^2$University of Science and Technology of China, Hefei 230026, China\\
  \small $^*$Corresponding authors: \texttt{zhiyuan.lu@ipp.ac.cn}, \texttt{gsxu@ipp.ac.cn}}
\date{}

\begin{document}
\maketitle

\begin{abstract}
Stellarators represent a leading pathway to steady-state, disruption-free magnetic fusion energy, yet their intrinsically three-dimensional magnetic fields induce uncompensated radial drifts of trapped particles, which exacerbates neoclassical transport and fast-ion losses. Omnigenity, the condition that the second adiabatic invariant \(J\) of trapped particles is invariant along magnetic field lines, is the fundamental orbit-physics principle that eliminates this net radial drift, but direct gradient-based optimization of its defining condition has long been intractable due to moving bounce boundaries, non-smooth cutoff singularities, and topological transitions in trapped-branch structure. Existing state-of-the-art approaches therefore rely on geometric or symmetry-based proxies derived from the Cary--Shasharina transformation, rather than targeting the trapped-particle action directly. Here we introduce a differentiable action-residual formulation that enables direct, gradient-based optimization of \(J\)-invariance within the DESC stellarator equilibrium and optimization framework. We systematically resolve the core numerical pathologies through normalized pitch sampling, softplus-regularized action integrands, log-sum-exp smooth extremum estimators, and a soft-connectivity regularizer that enforces physically interpretable single-branch bounce orbits. Applying this approach, we design a compact four-field-period stellarator with an aspect ratio of 4.3 and near-zero alpha-particle losses, alongside a balanced configuration that combines finite-\(\beta\) ideal-ballooning stability, reduced neoclassical transport, and compatibility with a practical modular coil set. Independent validation using exact hard-cutoff bounce action calculations confirms superior \(J\)-invariance relative to the Wendelstein 7-X reference across most pitch values. This work establishes a direct orbit-physics optimization paradigm for stellarators, enabling rigorous trapped-particle constraints to be integrated into multi-objective design of next-step fusion reactors.
\end{abstract}


\hypertarget{introduction}{%
\section{Introduction}\label{introduction}}

Magnetic confinement fusion holds the promise of safe, abundant, and carbon-free baseload energy, and stellarators stand out as one of the most promising reactor concepts owing to their inherent capability for steady-state, disruption-free operation. Unlike axisymmetric tokamaks, which rely on large plasma currents to confine the plasma, stellarators achieve confinement entirely through externally generated three-dimensional magnetic fields. This absence of a plasma current eliminates major risk drivers for reactor-scale devices, but it also breaks the continuous toroidal symmetry that suppresses neoclassical transport in tokamaks. In three-dimensional configurations, trapped particles exhibit uncompensated radial drifts that can substantially degrade energy confinement and cause significant fast-ion losses, making the mitigation of these drifts a central challenge in stellarator design.

Omnigenity, the condition that the second adiabatic invariant \(J\) of trapped particles is invariant along magnetic field lines, is the fundamental orbit-physics principle that eliminates net radial drift and can bring neoclassical transport down to tokamak-comparable levels. As the most general collisionless confinement condition, omnigenity subsumes quasisymmetry as a special case and admits several distinct geometric implementations, including quasi-isodynamic (QI) and quasi-helical (QH) configurations. Its physical significance has been experimentally validated on the Wendelstein 7-X device~\cite{beidler2021demonstration}. It has since become a cornerstone of modern stellarator optimization~\cite{duff2025suppressing,goodman2024quasi,garciaregana2025reduced}. This has motivated broad theoretical and computational studies of quasi-symmetric and quasi-isodynamic configurations~\cite{bindel2023direct,sanchez2023quasi,bonofiglo2025fast}. Recent work has further integrated these constraints into practical design frameworks~\cite{landreman2022optimization,alonso2022physics,jorge2023single}. Single-stage plasma--coil optimization has also been explored within this paradigm~\cite{wechsung2022single}. Despite its central importance, direct gradient-based optimization of this defining condition has long been computationally intractable, hampered by moving bounce boundaries, non-smooth square-root singularities at turning points, and topological changes in trapped-branch structure. As a result, state-of-the-art approaches have relied on geometric or symmetry-based proxies rooted in the Cary--Shasharina transformation, rather than targeting the trapped-particle action directly. In this work, we develop a fully differentiable action-residual formulation that enables direct, gradient-based optimization of \(J\)-invariance, and demonstrate its utility by designing high-performance stellarator configurations that balance exceptional orbit confinement with practical engineering feasibility.

The dominant strategies for omnigenity optimization are both rooted in the Cary--Shasharina (C--S) transformation~\cite{cary1997helical,cary1997omnigenity}, which recasts omnigenity as a geometric condition on constant-field-strength contours in a field-line-aligned coordinate system. One class of approaches constructs exactly omnigenous target field profiles from parameterized well shapes and coordinate mappings, then minimizes the pointwise mismatch between the equilibrium field and the target~\cite{dudt2024general}. This target-field paradigm has yielded high-confinement configurations, but it requires a priori specification of the magnetic well topology and helicity, and its objective remains a geometric proxy rather than a direct orbit-physics constraint. The second major strategy, exemplified by the OOPS framework~\cite{liu2025hidden}, optimizes hidden-symmetry coordinates to suppress asymmetric Fourier modes of the field strength. While powerful and flexible, this approach likewise enforces symmetry-based rather than action-based optimality, and its interpretation in terms of individual trapped-particle orbits remains indirect. To date, no gradient-based method has been able to directly optimize the defining condition of omnigenity, field-line invariance of \(J\), due to fundamental numerical pathologies: moving bounce boundaries that violate standard Leibniz differentiation, non-smooth cutoffs at turning points, and topological transitions between single- and multi-well trapped regions that introduce spurious gradient signals.

In this work, we overcome these barriers by developing a fully differentiable action-residual formulation that enables direct gradient-based optimization of \(J\)-invariance within the DESC stellarator equilibrium framework~\cite{dudt2020desc}. We address the core numerical pathologies through normalized pitch sampling, softplus-regularized action integrands, log-sum-exp smooth extrema, and a soft-connectivity regularizer. We demonstrate the method by designing two four-field-period stellarator configurations spanning a practical design trade-off between compactness, confinement, stability, and engineering feasibility. Crucially, unlike existing approaches that depend on a C--S target or mapping, our objective is computed directly from the equilibrium field without a priori specification of helicity or well topology. The remainder of the paper presents the theoretical formulation, numerical implementation, optimization results, and validation in detail.

\hypertarget{physical-formulation}{%
\section{Orbit-Based Formulation of Omnigenity}\label{physical-formulation}}

\hypertarget{boozer-field-line-label}{%
\subsection{Boozer Field-Line Label}\label{boozer-field-line-label}}

On a nested flux surface, let \((\rho,\theta_B,\zeta_B)\) denote Boozer coordinates. The angles \(\theta_B\) and \(\zeta_B\) are the Boozer poloidal and toroidal angles, and magnetic field lines are straight in these coordinates. A field line can be labeled by

\begin{equation}
\alpha = \theta_B - \iota(\rho)\zeta_B ,
\end{equation}
where \(\iota(\rho)\) is the rotational transform. Along a field line at fixed \(\rho\) and \(\alpha\),

\begin{equation}
\theta_B = \alpha + \iota(\rho)\zeta_B .
\end{equation}

For a smooth quantity evaluated at fixed \(\rho\) and fixed \(\zeta_B\), the pointwise chain-rule statement is

\begin{equation}
\left.
\frac{\partial Q(\rho,\theta_B(\alpha,
  \zeta_B),\zeta_B)}{\partial\alpha}
\right|_{\rho,\zeta_B}
=
\left.
\frac{\partial Q}{\partial\theta_B}
\right|_{\rho,\zeta_B}
\end{equation}

Since \(\theta_B=\alpha+\iota(\rho)\zeta_B\) and \(\iota\) depends only on \(\rho\), at fixed \(\rho\) and \(\zeta_B\) we have \(\partial\theta_B/\partial\alpha=1\), which gives the chain-rule result above. Here \(Q\) denotes a smooth pointwise field quantity, such as \(|B|\) or a Boozer geometric weight. We use this relation only after replacing the hard bounce action by the fixed-interval, smooth action proxy \(\widetilde{J}\).

\begin{figure}[!htbp]
\centering
\includegraphics[width=0.92\linewidth]{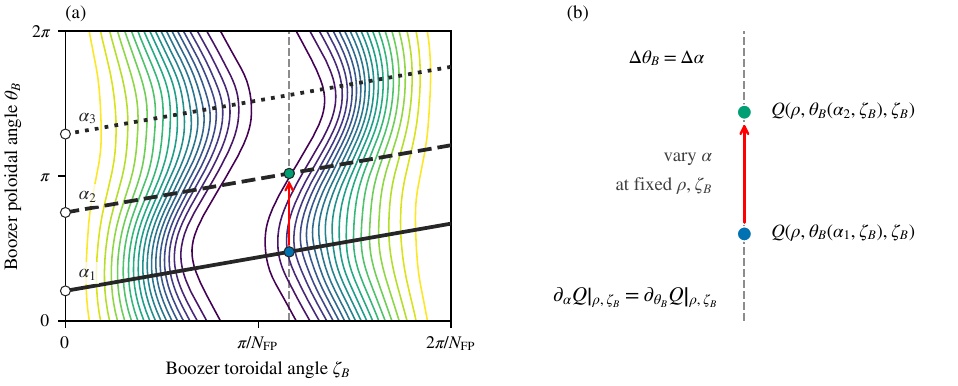}
\caption{Field-line labels and local \(\alpha\) displacement in Boozer coordinates at \(\rho=1\). Panel (a) overlays the field lines on normalized \(|B|\) for an omnigenous configuration. Panel (b) illustrates the local \(\alpha\) displacement at fixed toroidal angle \(\zeta_B\).}
\label{fig:boozer-label}
\end{figure}

Figure~\ref{fig:boozer-label} indicates the geometric meaning of the derivative used below. Moving along a field line holds \(\alpha\) fixed and changes \(\zeta_B\), whereas \(\partial/\partial\alpha\) compares neighboring field lines at the same \((\rho,\zeta_B)\). Thus the derivative in the action residual is not a derivative along the field line: after the bounce action is replaced by a smooth fixed-domain integral, it can be evaluated as the Boozer-poloidal derivative of the pointwise integrand.

\hypertarget{second-adiabatic-invariant}{%
\subsection{Second Adiabatic Invariant}\label{second-adiabatic-invariant}}

For a trapped particle with energy \(E\), magnetic moment \(\mu\), and mass \(m\), the parallel velocity satisfies

\begin{equation}
v_\parallel^2
=
\frac{2}{m}\left(E-\mu B\right).
\end{equation}

The second adiabatic invariant is the bounce action

\begin{equation}
J = \oint m v_\parallel\,\dd l ,
\label{eq:second-adiabatic-invariant}
\end{equation}
where the integral is taken over one complete bounce period. With \(B^{*}=E/\mu\), the part of a field line accessible to the particle is defined by \(B<B^{*}\). Up to constants that do not affect a zero-target residual, the action contains the factor

\begin{equation}
\sqrt{1-\frac{B}{B^{*}}} .
\end{equation}

In an omnigenous field, \(J\) depends on the flux surface and particle constants, but not on \(\alpha\). At fixed \(\rho\) and pitch,

\begin{equation}
J(\rho,\alpha,t)=J(\rho,t),
\qquad
\frac{\partial J(\rho,\alpha,t)}{\partial\alpha}=0 .
\end{equation}

This is the orbit condition targeted in the optimization.

\hypertarget{normalized-pitch-sampling}{%
\subsection{Normalized Pitch Sampling}\label{normalized-pitch-sampling}}

A fixed dimensional \(B^{*}\) would correspond to a shifting population of trapped particles as the equilibrium evolves, which would introduce spurious gradients and distort the optimization process. As \(B_{\min}\) and \(B_{\max}\) evolve, the same \(B^{*}\) would drift through the trapping range and could eventually satisfy \(B^{*}\leq B_{\min}\) or reach and cross \(B_{\max}\), for which there is no finite accessible region, where the orbit approaches the trapped--passing separatrix and then becomes passing. In practice, the \(B_{\min}\) area (in Quasi-Isodynamic case, around \(\zeta_B=\frac{\pi}{N_{FP}}\)) would be pushed far below preset \(B^{*}\), since \(B_{\min}\) area is actually difficult to optimize. Such an ordinary solution is not what we wanted, so we use a normalized pitch coordinate \(t\in(0,1)\) and define

\begin{equation}
B^{*}(\rho,t)
=
(1-t)B_{\min}(\rho)+tB_{\max}(\rho).
\end{equation}

Here \(B_{\min}\) and \(B_{\max}\) are the minimum and maximum field strength on the sampled flux surface. Thus \(t=0\) and \(t=1\) denote the local low- and high-field ends of the trapping range, while each intermediate \(t\) denotes a fixed fractional position within that range. Under an overall rescaling \(B\mapsto cB\), the pitch reference scales as \(B^{*}\mapsto cB^{*}\), leaving the ratio \(B/B^{*}\) in the action integrand unchanged. A fixed \(t\) grid therefore follows the instantaneous trapping range and samples comparable trapped-orbit classes across surfaces and optimization iterations. This prevents samples from entering or leaving the trapped range solely because the field scale changes, so variations in the residual primarily reflect changes in the magnetic geometry rather than changes in the sampled particle population. When \(B_{\min}\) and \(B_{\max}\) are evaluated with the smooth extrema introduced in Section~\ref{smooth-extrema}, \(B^{*}\) also varies smoothly with the equilibrium degrees of freedom. Combined with the smooth cutoff of Section~\ref{smoothed-action-integrand}, this preserves a continuous differentiable path through \(B/B^{*}\) to the action residual and avoids abrupt or poorly scaled gradients as the equilibrium evolves. Normalization therefore supplies sampling consistency and field-scale invariance; the smooth extrema and cutoff supply differentiability.

\hypertarget{relation-to-existing-approaches}{%
\subsection{Relation to Existing Approaches}\label{sec:existing}}

The present method differs fundamentally from existing omnigenity optimization strategies in the quantity being optimized and in its relationship to the Cary-Shasharina (C-S) construction. The C-S transformation \cite{cary1997helical,cary1997omnigenity} is the unifying mathematical framework behind both major approaches. It introduces coordinates \((\alpha,\eta)\) on a flux surface: \(\alpha\) is the Clebsch field-line label, and \(\eta\in[-\pi/2,\pi/2]\) labels constant-\(B\) contours from the minimum (\(\eta=0\)) to the maximum (\(\eta=\pm\pi/2\)). A prescribed well shape \(B(\eta)\), monotonic, with vanishing derivatives at the endpoints, sets the magnetic-field profile along each contour. The helicity \((M,N)\) of the \(B\)-contour topology, whether poloidal (\(M=0\)), toroidal (\(N=0\)), or helical (\(M,N\neq0\)), is encoded in the mapping from \((\alpha,\eta)\) back to Boozer angles \((\theta_B,\zeta_B)\). The core C-S omnigenity condition is \(\partial\Delta\zeta_B/\partial\alpha=0\), where \(\Delta\zeta_B\) is the toroidal angular distance along a field line between consecutive equal-\(B\) bounce points. As a geometric byproduct, this straightens the \(B_{\max}\) contour in the \((\alpha,\eta)\) plane, though the straight-contour property is necessary but not sufficient for omnigenity. The well-known C-S no-go theorem, that only quasi-symmetric fields can satisfy omnigenity analytically, explains why all practical omnigenity optimization is inherently numerical.

The two existing strategies use the C-S construction in complementary ways. In the target-field approach of Dudt et al.~\cite{dudt2024general}, which generalizes the C-S construction to arbitrary helicity following Landreman and Catto~\cite{landreman2012omnigenity}, an exactly omnigenous \(|B|\) target is constructed from a parameterized well shape \(B(\eta)\) and Fourier coefficients \(x_{lmn}\) of the C-S coordinate mapping. The optimizer varies both the equilibrium parameters and the target-field parameters, namely the well-shape knots \(B_{ij}\) and the mapping coefficients \(x_{lmn}\), to minimize the pointwise mismatch \(|B_{\mathrm{eq}}-B_{\mathrm{target}}|^2\). This approach treats the C-S construction as a \emph{target generator}: the desired omnigenous field is fully specified before the equilibrium optimization begins. The OOPS framework~\cite{liu2025hidden} takes a fundamentally different approach: it treats the C-S mapping as a \emph{symmetry-detecting coordinate basis}. A homeomorphic (continuous, invertible) mapping \(\mathcal{M}: (\alpha,\eta) \rightarrow (\theta_B,\zeta_B)\) is parameterized to straighten \(|B|\) contours, and omnigenity is enforced by suppressing the asymmetric Fourier modes \(B_{m,n}\) (\(m \neq 0\)) of \(B\) evaluated in the transformed coordinates. The objective \(f_{\mathrm{symm}} = \sum_{m\neq 0} (B_{m,n}/B_{0,0})^2\) has the same algebraic form as quasi-symmetry optimization but is evaluated after applying \(\mathcal{M}^{-1}\). Different hidden symmetries, including quasisymmetry, omnigenity, pseudosymmetry, and piecewise omnigenity \cite{velasco2024piecewise}, correspond to different constraints on the mapping parameters (e.g., pseudosymmetry retains the no-locally-closed-contours condition while relaxing \(\partial\Delta\zeta_B/\partial\alpha=0\)). Both strategies are powerful because they encode the C-S structure through a parameterized target or mapping, but their objectives are geometric (\(|B|\) mismatch) or symmetry-based (Fourier-mode amplitudes), rather than directly tied to the trapped-particle orbit action \(\partial J/\partial\alpha=0\).

In the action-residual approach, the C-S construction is not used. The optimizer does not need a prescribed helicity, well-shape parameterization, constructed \(|B|\) template, or homeomorphic mapping. It instead penalizes the field-line variation of the action computed directly from the current equilibrium through the Boozer spectral representation. This removes the need for a priori parameterization choices and keeps the objective directly tied to trapped-particle orbit physics. However, giving up the structured C-S target or mapping also requires careful regularization: the residual must remain smooth, properly scaled, and physically interpretable when the trapped region changes topology.

\hypertarget{differentiable-action-residual}{%
\section{Differentiable Action-Residual Formulation}\label{differentiable-action-residual}}

\hypertarget{smoothed-action-integrand}{%
\subsection{Smoothed Action Integrand}\label{smoothed-action-integrand}}

Starting from the second adiabatic invariant in Eq.~\eqref{eq:second-adiabatic-invariant}, and using \(B^{*}=E/\mu\), the parallel velocity can be written as

\begin{equation}
v_\parallel
=
\sqrt{\frac{2\mu B^{*}}{m}}
\sqrt{1-\frac{B}{B^{*}}}.
\end{equation}

Between two bounce points \(l_{b1}\) and \(l_{b2}\), where \(B=B^{*}\), one complete back-and-forth bounce therefore gives

\begin{equation}
J
=
2\sqrt{2m\mu B^{*}}
\int_{l_{b1}}^{l_{b2}}
\sqrt{1-\frac{B}{B^{*}}}
\,\dd l .
\end{equation}

The square root is physically defined only in the accessible region \(E-\mu B\geq0\), or equivalently \(B\leq B^{*}\). When the bounce integral is represented on a fixed grid that also contains inaccessible points, this restriction is expressed by taking the positive part of its argument. The hard action integrand is therefore

\begin{equation}
\sqrt{\max\left(1-\frac{B}{B^{*}},0\right)} .
\end{equation}

Thus, the hard cutoff is not an additional orbit model; it is the fixed-domain representation of the physical condition \(v_\parallel^2\geq0\). However, the maximum is non-smooth at the bounce points, and the derivative of the square root is singular as the accessible side approaches \(B=B^{*}\). To retain the physical meaning of the trapped-particle action while making its numerical representation differentiable, we replace the positive-part operation with a softplus approximation,

\begin{equation}
\operatorname{sp}_\epsilon(x)=
\epsilon \log\left(1+\exp(x/\epsilon)\right),
\end{equation}
and use

\begin{equation}
\sqrt{\operatorname{sp}_\epsilon\left(1-\frac{B}{B^{*}}\right)}
\end{equation}
inside the action integrand. The argument of \(\operatorname{sp}_\epsilon\) is dimensionless, so we choose \(\epsilon = 0.01\) (corresponding to a softplus sharpness of \(\beta = 1/\epsilon = 100\)) for the normalized cutoff argument; equivalently, the corresponding field-unit smoothing scale is 1\% of \(B_{\max}-B_{\min}\) on the flux surface. This value is sufficiently small so that the hard-action validation is insensitive to further reduction, while remaining large enough to remove singular derivatives at the bounce points.

The quantity \(\widetilde{J}\) below is an action proxy, not the exact second adiabatic invariant: it regularizes the cutoff and uses a fixed integration interval. It is introduced so that its field-line derivative can be used during optimization, while the exact hard-cutoff action is evaluated independently for validation. The smoothed action proxy is

\begin{equation}
\widetilde{J}(\rho,\alpha,t)=
\int_{0}^{2\pi/N_{\mathrm{FP}}}
\sqrt{\operatorname{sp}_\epsilon\!\left(1-\frac{B(\rho,\alpha+\iota\zeta_B,\zeta_B)}{B^{*}(\rho,t)}\right)}
\;
\sqrt{g}_{\mathrm{Boozer}}
\;
\dd\zeta_B ,
\end{equation}
where \(\sqrt{g}_{\mathrm{Boozer}}\) is the Jacobian of the Boozer-to-Cartesian coordinate transformation, \(N_{\mathrm{FP}}\) is the number of field periods, the integration runs over one Boozer toroidal field period \(2\pi/N_{\mathrm{FP}}\), and \(\theta_B=\alpha+\iota\zeta_B\) traces the field line.

The fixed-interval proxy is essential beyond the removal of the square-root singularity. In the original definition, the bounce limits \(\zeta_{b1}(\alpha)\) and \(\zeta_{b2}(\alpha)\) depend on the field-line label and on the optimization variables. Differentiating such an integral requires the Leibniz rule,

\begin{equation}
\frac{\dd}{\dd\alpha}\int_{a(\alpha)}^{b(\alpha)} f\,\dd\zeta
= \int_{a}^{b} \frac{\partial f}{\partial\alpha}\,\dd\zeta
+ f\big|_{b}\,\frac{db}{d\alpha}
- f\big|_{a}\,\frac{da}{d\alpha},
\end{equation}

The moving limits introduce boundary terms whose locations are defined implicitly by \(B(\zeta)=B^{*}\), making gradients with respect to \(\alpha\) and the equilibrium parameters sensitive to representation and discretization. Extending the smoothed integrand to the fixed domain \([0,2\pi/N_{\mathrm{FP}}]\) removes these moving-boundary terms, so the \(\alpha\)-derivative acts only on a smooth integrand.

\hypertarget{analytical-alpha-derivative}{%
\subsection{Analytical Field-Line Derivative of the Action}\label{analytical-alpha-derivative}}

The residual is obtained by differentiating the smoothed action proxy \(\widetilde{J}\). Because the integration domain \([0, 2\pi/N_{\mathrm{FP}}]\) is fixed (does not depend on \(\alpha\)) and the integrand is everywhere smooth, the \(\alpha\)-derivative commutes with the \(\zeta_B\)-integral:

\begin{equation}
\frac{\partial \widetilde{J}}{\partial\alpha}
=
\int_{0}^{2\pi/N_{\mathrm{FP}}}
\frac{\partial}{\partial\alpha}
\left[
\sqrt{\operatorname{sp}_\epsilon\!\left(1-\frac{B}{B^{*}}\right)}
\;
\sqrt{g}_{\mathrm{Boozer}}
\right]
\dd\zeta_B .
\end{equation}

With \(\rho\), \(t\), \(\zeta_B\), and \(B^{*}(\rho,t)\) held fixed, \(\alpha\) enters the integrand only through the field-line parameterization \(\theta_B = \alpha + \iota\zeta_B\). The \(\alpha\)-derivative therefore reduces to a \(\theta_B\)-derivative:

\begin{equation}
\frac{\partial \widetilde{J}}{\partial\alpha}
=
\int_{0}^{2\pi/N_{\mathrm{FP}}}
\frac{\partial}{\partial\theta_B}
\left[
\sqrt{\operatorname{sp}_\epsilon\!\left(1-\frac{B}{B^{*}}\right)}
\;
\sqrt{g}_{\mathrm{Boozer}}
\right].
\dd\zeta_B .
\end{equation}

The \(\theta_B\)-derivative is then evaluated by automatic differentiation through the Boozer spectral representation, which supplies \(\partial B/\partial\theta_B\) and \(\partial\sqrt{g}_{\mathrm{Boozer}}/\partial\theta_B\) analytically from the Fourier coefficients, giving AD a continuous computational graph without any finite-difference step.

Two further advantages follow from this formulation. First, it avoids finite-difference estimation of \(\partial J/\partial\alpha\) entirely. If one were to compute \(J(\alpha)\) on a discrete \(\alpha\)-grid and then apply finite differences, each \(J\) evaluation would itself involve a numerical quadrature with its own discretization error; the differencing step would then amplify both the quadrature noise and the truncation error of the finite-difference stencil. Moreover, when the trapped region splits into multiple wells, the global \(J(\alpha)\) may switch between branches at different \(\alpha\) values, making the finite-difference value physically meaningless. In contrast, the analytical derivative is obtained directly through the AD computational graph, and the single-well regime is enforced separately by the soft-connectivity penalty (Section~\ref{sec:soft-connectivity}). Second, the \(\zeta_B\)-integral acts as a smoothing operation on the pointwise \(\theta_B\)-derivatives. Near the bounce points, even the softplus-smoothed integrand can have steep gradients that would dominate a pointwise residual. Integrating these derivatives over the full field period averages out local oscillations and yields the global \(\alpha\)-variation of the action, which is precisely the quantity that omnigenity requires to vanish, without being sensitive to where along the field line the variation occurs.

\hypertarget{smooth-extrema}{%
\subsection{Differentiable Evaluation of Bounce Extrema}\label{smooth-extrema}}

The normalized pitch definition depends on \(B_{\min}\) and \(B_{\max}\). Hard extrema are not differentiable when the grid point that sets the extremum changes. Because \(B^{*}\) enters every action integrand, this would break the complete gradient chain from equilibrium parameters to \(f_J\), even if the action integrand itself were smooth. We therefore approximate the extrema by logsumexp smooth extrema:

\begin{equation}
B_{\max}
\approx
\sigma_B \log\sum_q \exp(B_q/\sigma_B),
\end{equation}
\begin{equation}
B_{\min}
\approx
-\sigma_B \log\sum_q \exp(-B_q/\sigma_B),
\end{equation}
where \(q\) indexes the \(N\) sampled points on the flux surface. In our implementation, the smoothing scale is adaptively set as \(\sigma_B = \tau_{\mathrm{rel}} (B_{\max}-B_{\min}) / \log N\) with \(\tau_{\mathrm{rel}} = 0.1\), which ensures that the relative overshoot of the smooth extrema is scaled to 10\% of the field variation \(B_{\max}-B_{\min}\) independent of grid resolution. This keeps the approximation stable across surfaces and preserves a differentiable path through the normalized pitch. Smooth extrema address differentiability only; they do not ensure that the allowed region \(B<B^{*}\) is a single trapped branch. That separate branch-consistency issue is handled by the soft-connectivity penalty below.

Combining the smooth extrema with the analytical \(\alpha\)-derivative from Section~\ref{analytical-alpha-derivative}, the total \(J\)-invariance residual on a flux surface \(\rho\) is evaluated on a discrete grid of \(N_\alpha\) field lines \(\{\alpha_i\}\), \(N_t\) pitch values \(\{t_j\}\), and \(N_\zeta\) toroidal samples \(\{\zeta_k\}\) per field period:

\begin{equation}
f_J(\rho)
=
\frac{1}{N_\alpha N_t N_\zeta}
\sum_{i=1}^{N_\alpha}
\sum_{j=1}^{N_t}
\sum_{k=1}^{N_\zeta}
\left|
\frac{\partial}{\partial\theta_B}
\left[
\sqrt{
\operatorname{sp}_\epsilon\!\left(
1-\frac{B(\rho,\alpha_i+\iota\zeta_k,\zeta_k)}
{B^{*}(\rho,t_j)}
\right)
}
\,
\sqrt{g}_{\mathrm{Boozer}}
\right]
\right|^2,
\end{equation}
where the pitch reference \(B^{*}(\rho,t_j) = (1-t_j)\widetilde{B}_{\min}(\rho) + t_j\widetilde{B}_{\max}(\rho)\) uses the logsumexp smooth extrema \(\widetilde{B}_{\min}\) and \(\widetilde{B}_{\max}\). The \(\theta_B\)-derivative at each grid point is obtained by automatic differentiation through the Boozer spectral representation. This discrete form makes the role of smooth extrema explicit: the approximations \(\widetilde{B}_{\min}\) and \(\widetilde{B}_{\max}\) enter the residual solely through \(B^{*}\), which appears in the denominator of the softplus argument \(1-B/B^{*}\) at every grid point of the triple sum. With hard extrema, a change in which grid point supplies \(B_{\min}\) or \(B_{\max}\) would produce a discontinuous jump in \(B^{*}\), breaking the differentiability of \(f_J\) with respect to the equilibrium parameters. Replacing them by the logsumexp operation, which is everywhere smooth, keeps \(B^{*}\) and therefore the entire integrand differentiable, closing the chain from the equilibrium degrees of freedom through the pitch normalization to the \(J\)-invariance objective.

\begin{figure}[!htbp]
\centering
\includegraphics[width=0.98\linewidth]{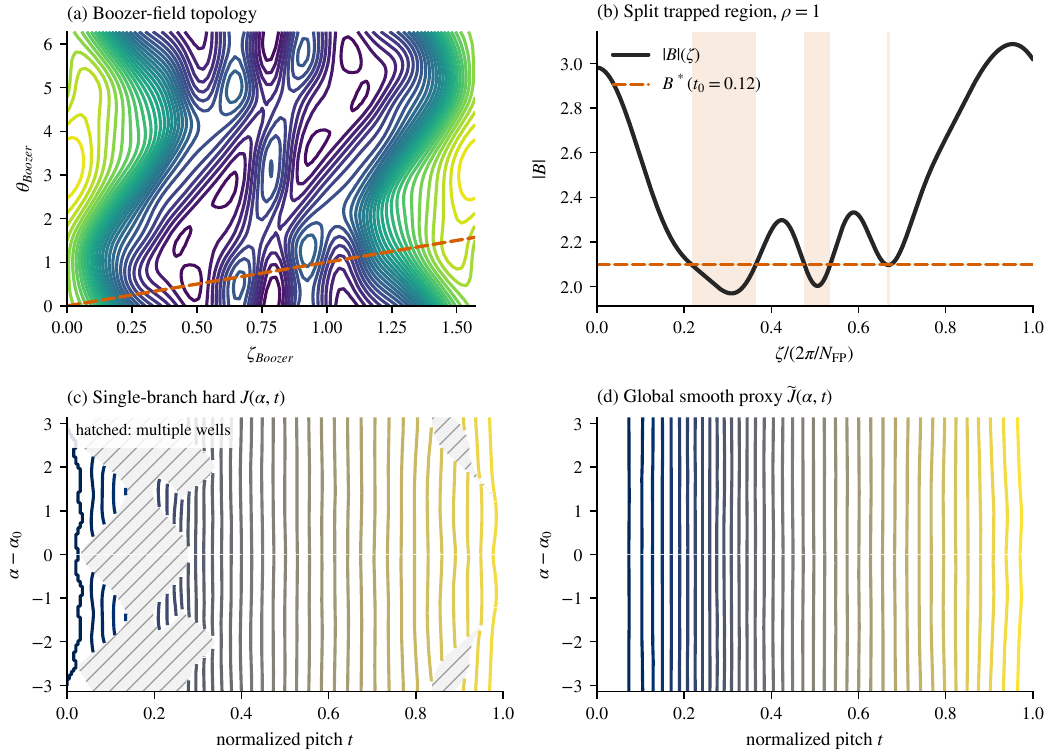}
\caption{Action diagnostics for the SC-off equilibrium at \(\rho=1\). (a) Boozer-surface \(|B|\) contours and the representative field line used in (b). (b) Field strength over one field period; the dashed threshold \(B^{*}\) produces several disconnected trapped intervals, shown by the shaded regions. (c) Single-branch hard-cutoff action, with multi-well samples hatched and excluded. (d) Fixed-domain smooth action proxy evaluated over the full sample grid. In panels (c) and (d), the color scale is the normalized bounce action \(J/J_{\max}\), and the vertical coordinate is \(\alpha-\alpha_0\).}
\label{fig:1}
\end{figure}

Figure~\ref{fig:1} illustrates the numerical regularization and its topological limitation for the soft-connectivity-disabled (SC-off) equilibrium at \(\rho=1\). Here, SC-off denotes an optimization in which the action-residual objective is retained but the soft-connectivity regularization is omitted. The hard action omits samples with split magnetic wells because a single global value of \(J\) does not then represent one bounce branch. The fixed-domain smooth proxy remains defined on the full sample grid, which is useful for optimization but does not itself control the trapped-region topology.

\hypertarget{objective-functions}{%
\section{Optimization Framework}\label{objective-functions}}

\hypertarget{soft-connectivity}{%
\subsection{Soft Connectivity Regularization}\label{sec:soft-connectivity}}

\hypertarget{branch-ambiguity-of-a-global-action}{%
\subsubsection{Motivation and Physical Interpretation}\label{sec:branch-ambiguity}}

The global action residual has a clear orbit meaning when the allowed region \(B<B^{*}\) is one connected interval along the sampled field line. In that case, the smoothed integral approximates one bounce branch, and reducing its \(\alpha\) variation directly reduces the splitting of that branch. Smooth action regularization addresses differentiability; it does not by itself ensure this branch consistency.

The interpretation changes when the same pitch produces several disconnected intervals. A global integral over \(\zeta_B\) then adds contributions from multiple bounce branches. In the multi-well regime, a small global \(\alpha\)-variation of \(J\) can therefore arise from cancellation of variations across disconnected branches, rather than genuine omnigenity of each individual bounce orbit. The soft-connectivity penalty suppresses this spurious optimization by enforcing a single-well topology, ensuring that the residual \(f_J\) has a clear physical interpretation. This is the structural ambiguity in any global action objective that does not resolve bounce branches.

A fully branch-resolved objective would have to identify each trapped interval, track it during optimization, and handle the creation, merger, and disappearance of intervals. Those events are non-smooth and difficult to combine with AD. We therefore restrict the optimization to configurations where the sampled trapped region remains single-well. This is a modeling choice that gives the global residual a single-branch interpretation, not a requirement that all omnigenous fields be single-well. Figure~\ref{fig:1}(b) shows the failure mode directly: for a fixed pitch, one horizontal \(B^{*}\) level can divide the same field-line trace into several disconnected trapped intervals.

\hypertarget{soft-connectivity-penalty}{%
\subsubsection{Formulation of the Soft Connectivity Objective}\label{soft-connectivity-penalty}}

The soft-connectivity penalty is a regularization term, not a second omnigenity objective: \(f_J\) measures \(J\)-invariance, whereas \(f_{\mathrm{SC}}\) enforces the branch consistency needed for that measurement to be physically interpretable. It uses spline curves in Boozer coordinates to divide each sampled field line into two parts: a left half where \(|B|\) should decrease toward the low-field point, and a right half where \(|B|\) should increase away from it. The division is made by two alpha-dependent curves represented by periodic cubic splines,

\begin{equation}
\zeta_{\min}(\alpha), \qquad\zeta_{\max}(\alpha),
\end{equation}

\begin{figure}[!htbp]
\centering
\includegraphics[width=0.98\linewidth]{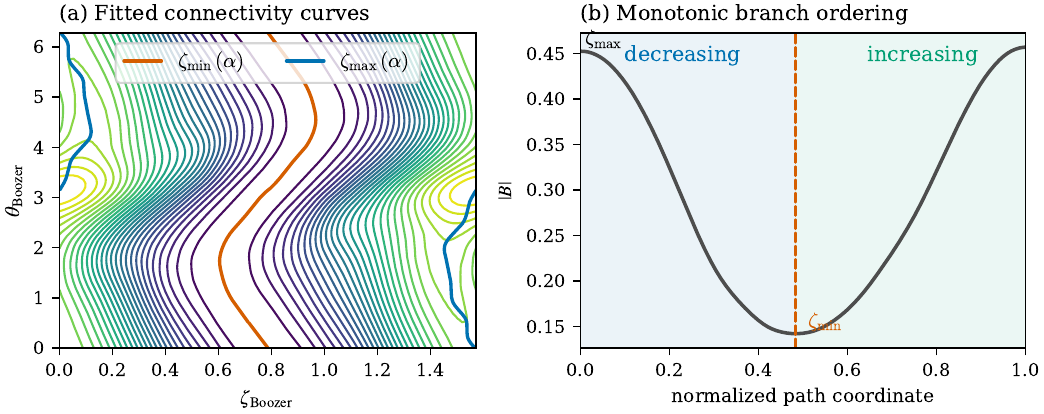}
\caption{Soft-connectivity construction for the compact configuration. The equilibrium is held fixed while the spline variables are refit. (a) Fitted low-field \(\zeta_{\min}(\alpha)\) and high-field \(\zeta_{\max}(\alpha)\) curves on the Boozer-coordinate plane. (b) Magnetic-field strength along the normalized field-line path, showing the required decreasing-then-increasing monotonic structure for a single well.}
\label{fig:2}
\end{figure}

Here \(\zeta_{\min}\) marks the low-field point on the field line, and \(\zeta_{\max}\) marks the high-field point near the field-period boundary. Cubic splines provide continuously differentiable branch boundaries that can be optimized with AD while remaining flexible enough to follow the field-line extrema. The fitted curves provide a direct diagnostic of whether the assumed one-well ordering is consistent with the Boozer-field structure. Figure~\ref{fig:2} shows this diagnostic on a fixed optimized equilibrium, with the spline variables refit only for the connectivity residual.

For a field line with label \(\alpha\), the sampled path starts from \(\zeta_{\max}(\alpha)\), passes through \(\zeta_{\min}(\alpha)\), and ends at the corresponding \(\zeta_{\max}\) in the next field period. It is clearer to write this path as two halves. With a normalized path coordinate \(\xi\in[0,1]\), the left half is

\begin{equation}
\zeta_L(\xi)
=
(1-\xi)\zeta_{\max}(\alpha)
+
\xi\zeta_{\min}(\alpha),
\qquad
\theta_L(\xi)=\alpha+\iota\zeta_L(\xi),
\end{equation}
and the right half is

\begin{equation}
\zeta_R(\xi)
=
(1-\xi)\zeta_{\min}(\alpha)
+
\xi\left[\zeta_{\max}(\alpha_{\mathrm{next}})+\frac{2\pi}{N_{\mathrm{FP}}}\right],
\qquad
\theta_R(\xi)=\alpha+\iota\zeta_R(\xi),
\end{equation}
where \(2\pi/N_{\mathrm{FP}}\) is one Boozer toroidal field period, and

\begin{equation}
\alpha_{\mathrm{next}}=\alpha+\iota\frac{2\pi}{N_{\mathrm{FP}}}.
\end{equation}

At the sampled nodes, the line derivative is evaluated as

\begin{equation}
\frac{\dd|B|}{\dd\zeta_{\mathrm{line}}}
=
\iota\frac{\partial |B|}{\partial\theta}
+
\frac{\partial |B|}{\partial\zeta}.
\end{equation}

The spline value \(\zeta_{\min}(\alpha)\) is used as a smooth switch between the two halves. With

\begin{equation}
s=\mathrm{sigmoid}\left[
k\left(\zeta-\zeta_{\min}\right)
\right],
\end{equation}
where \(k\) is \texttt{sigmoid\_sharpness}, the residual at each sampled point is

\begin{equation}
R_{\mathrm{SC}}
=
(1-s)\,\operatorname{softplus}
\left(
\frac{\dd|B|}{\dd\zeta_{\mathrm{line}}}
\right)
+
s\,\operatorname{softplus}
\left(
-\frac{\dd|B|}{\dd\zeta_{\mathrm{line}}}
\right).
\end{equation}

Thus \(R_{\mathrm{SC}}\) is positive when the field increases on the left branch or decreases on the right branch, both of which signal a wrong ordering or a split trapped region; it vanishes when the desired decrease-then-increase ordering is satisfied. The total soft-connectivity objective is the mean over all sampled field lines and pitch values,

\begin{equation}
f_{\mathrm{SC}}
=
\frac{1}{N_\alpha N_t N_\zeta}
\sum_{i,j,k}
R_{\mathrm{SC}}(\rho,\alpha_i,t_j,\zeta_k).
\end{equation}
\hypertarget{complete-objective-set}{%
\subsection{Complete Optimization Objective}\label{complete-objective-set}}

The optimization is performed in the DESC framework using a joint state \((x_{\mathrm{eq}}, x_{\mathrm{spline}})\), where \(x_{\mathrm{eq}}\) contains the equilibrium parameters and \(x_{\mathrm{spline}}\) contains the spline parameters defining the soft-connectivity reference curves. The MHD, aspect-ratio, elongation, rotational-transform, and magnetic-well terms use existing DESC objective functionality and impose equilibrium and engineering constraints. The two new orbit-physics terms are \(f_J\), which targets field-line invariance of the action proxy, and \(f_{\mathrm{SC}}\), which keeps that global action on a consistent trapped branch. With explicit objective weights, the complete optimization problem is

\begin{equation}
f = w_{\mathrm{MHD}}f_{\mathrm{MHD}} + w_{\mathrm{A}}f_{\mathrm{A}} + w_{\mathrm{E}}f_{\mathrm{E}} + w_{\iota,0}f_{\iota,0} + w_{\iota,1}f_{\iota,1} + w_J f_J + w_{\mathrm{SC}}f_{\mathrm{SC}} + w_{\mathrm{well}}f_{\mathrm{well}} .
\end{equation}

The residual blocks are:

\begin{itemize}
\tightlist
\item
  \(f_{\mathrm{MHD}}\): current-density residual, substitution for force balance in vacuum field.
\item
  \(f_{\mathrm{A}}\): aspect-ratio target.
\item
  \(f_{\mathrm{E}}\): elongation target.
\item
  \(f_{\iota,0}\) and \(f_{\iota,1}\): rotational-transform targets on an inner surface and on the boundary.
\item
  \(f_J\): alpha derivative of the smoothed action proxy for the second adiabatic invariant.
\item
  \(f_{\mathrm{SC}}\): soft-connectivity residual.
\item
  \(f_{\mathrm{well}}\): magnetic-well lower-bound residual.
\end{itemize}

\hypertarget{results}{%
\section{Results}\label{results}}

The results are organized around three applications of the formulation. A controlled SC-off/SC-on pair isolates the role of soft connectivity, a compact configuration establishes the confinement performance accessible when compactness and orbit quality are prioritized, and a balanced configuration tests compatibility with finite-pressure and coil-realizability. All alpha-particle loss fractions reported below were computed with SIMPLE~\cite{albert2020accelerated,albert2020symplectic} after each equilibrium was rescaled to the ARIES-CS device scale, with minor radius \(a=1.7\,\mathrm{m}\) and mean magnetic-field strength on the magnetic axis \(B_{00}=5.7\,\mathrm{T}\).

\subsection{Performance of Soft Connectivity}\label{sec:sc-performance}

The controlled pair differs only in whether soft connectivity objective is enforced. The resulting global properties remain closely matched, \((A,E,\iota_{\mathrm{edge}})=(10.0,4.95,1.00)\) for both. This close agreement isolates the confinement consequences of the different trapped-region topology.

\begin{figure}[!htbp]
\centering
\includegraphics[width=0.98\linewidth]{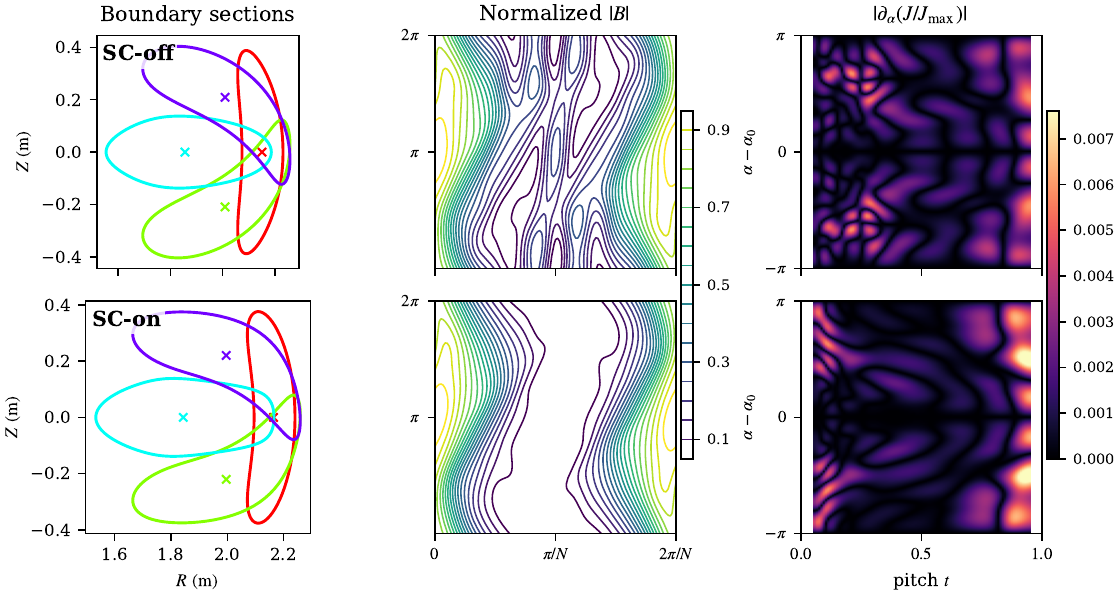}
\caption{Controlled soft-connectivity comparison at \(\rho=1\). Rows show SC-off and SC-on; columns show boundary sections, normalized Boozer-surface field strength, and \(\lvert\partial_\alpha(J/J_{\max})\rvert\) for the smooth action. The derivative color scale starts at zero and is adapted to the controlled pair.}
\label{fig:config-sc}
\end{figure}

\begin{figure}[!htbp]
\centering
\includegraphics[width=0.98\linewidth]{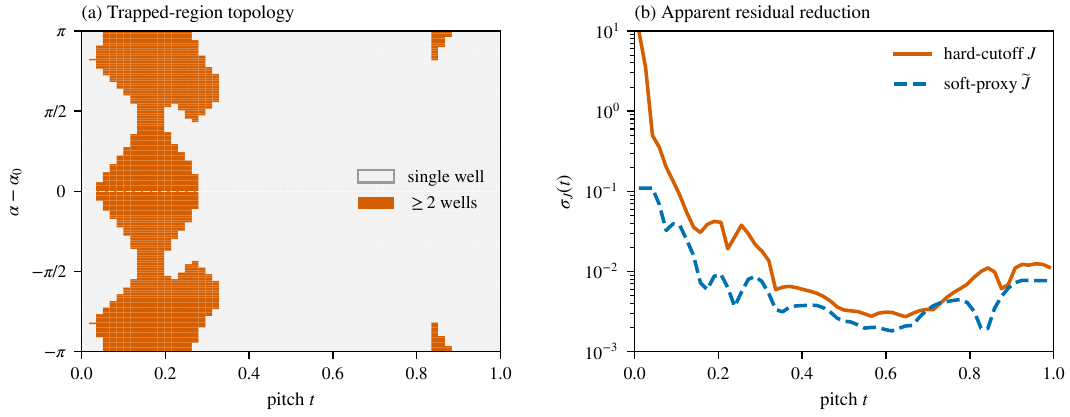}
\caption{Topology diagnostics for the SC-off equilibrium. (a) Regions in \((\alpha,t)\) where the allowed set \(B<B^{*}\) contains at least two disconnected wells. (b) Normalized field-line variation of the hard-cutoff action and smooth proxy on the same final equilibrium. The logarithmic scale exposes the proxy's under-estimation over the full pitch range.}
\label{fig:sc-topology}
\end{figure}

Figure~\ref{fig:config-sc} separates the topological role of soft connectivity from changes in global geometry. The boundary sections remain similar, but their Boozer-field structures differ qualitatively: SC-off contains fragmented low-field contours, whereas SC-on recovers a coherent single-well structure. Figure~\ref{fig:sc-topology} makes the physical consequence explicit. When a field line contains disconnected trapped intervals, the smooth action can show weak \(\alpha\)-variation while the individual trapped branches remain poorly aligned.

\begin{figure}[!htbp]
\centering
\includegraphics[width=0.98\linewidth]{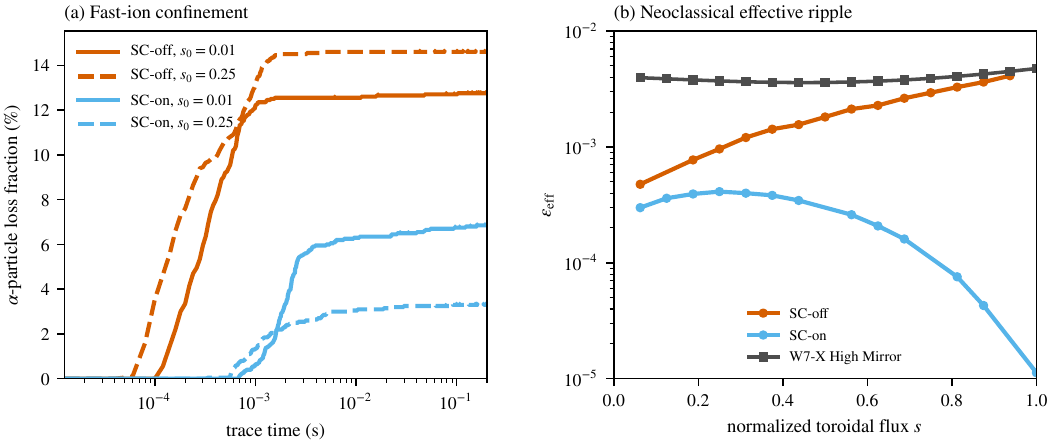}
\caption{Confinement diagnostics for the controlled pair. (a) Alpha-particle loss traces for every configuration--launch-surface combination; the legend gives both the configuration and \(s_0\) associated with each color--line-style pair. (b) Neoclassical effective ripple for SC-off and SC-on, with W7-X High Mirror as a reference. Unavailable radial samples are omitted.}
\label{fig:conf-sc}
\end{figure}

The independent confinement diagnostics in Fig.~\ref{fig:conf-sc} show the same controlled improvement. With 2000 alpha particles followed for 0.2~s at each launch radius, SC-off loses 12.75\% at \(s_0=0.01\) and 14.60\% at \(s_0=0.25\), whereas SC-on loses 6.85\% and 3.30\%. These changes correspond to reductions of 46.3\% and 77.4\%, respectively. At every radius where both effective-ripple profiles are available, SC-on also has lower \(\varepsilon_{\mathrm{eff}}\)~\cite{nemov1999evaluation}, with a median reduction of 77.8\%. W7-X High Mirror is shown as the common neoclassical reference.

\begin{table}[!htbp]
\centering
\caption{Geometric and confinement metrics for the controlled soft-connectivity comparison.}
\label{tab:sc-metrics}
\small
\begin{tabularx}{\linewidth}{@{}l*{5}{>{\centering\arraybackslash}X}@{}}
\toprule
Configuration & \(A\) & \(E\) & Edge well & \(\alpha\) loss, \(s_0=0.01\) & \(\alpha\) loss, \(s_0=0.25\) \\
\midrule
SC-off & 10.00 & 4.95 & \(-1.27\times10^{-2}\) & 12.75\% & 14.60\% \\
SC-on  & 10.00 & 4.95 & \(-1.57\times10^{-2}\) & 6.85\%  & 3.30\% \\
\bottomrule
\end{tabularx}
\end{table}

The controlled comparison therefore shows that soft connectivity is physically consequential rather than merely a numerical regularization. It restores a coherent trapped region and improves both fast-particle and neoclassical confinement without appreciably changing the global geometric properties.
\FloatBarrier

\subsection{Compact Configuration}\label{sec:compact-configuration}

\begin{figure}[!htbp]
\centering
\includegraphics[width=0.98\linewidth]{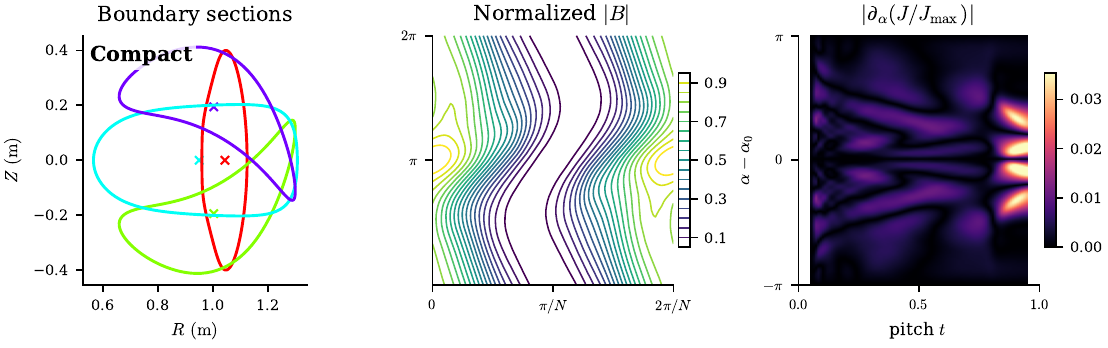}
\caption{Compact configuration at \(\rho=1\): boundary sections, normalized Boozer-surface field strength, and \(\lvert\partial_\alpha(J/J_{\max})\rvert\) for the smooth action. The derivative color scale starts at zero and is adapted to this configuration.}
\label{fig:config-compact}
\end{figure}

The compact configuration has four field periods, aspect ratio of \(A=4.30\), elongation of \(E=5.00\), and a negative magnetic-well. Its boundary and field-strength structure are shown in Fig.~\ref{fig:config-compact}. The negative magnetic well indicates unfavorable vacuum MHD stability properties, representing a key trade-off between compactness and stability that motivates the balanced configuration.

\begin{figure}[!htbp]
\centering
\includegraphics[width=0.98\linewidth]{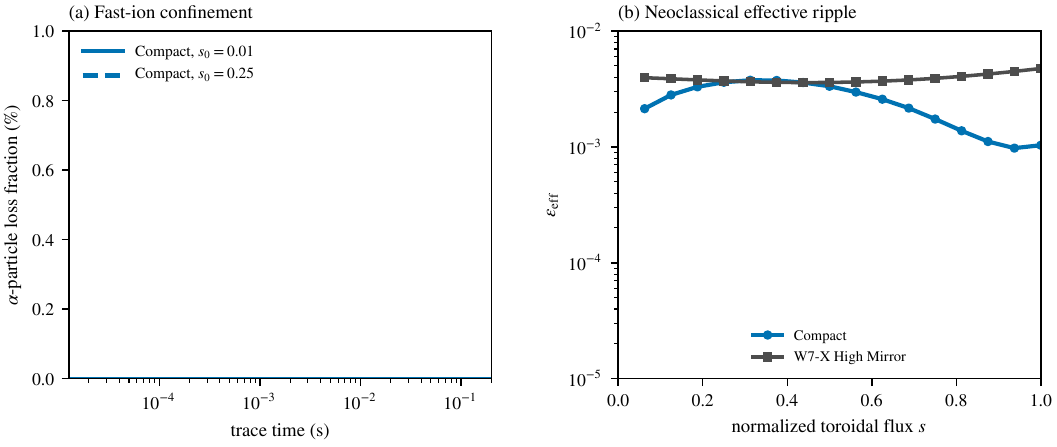}
\caption{Confinement diagnostics for the compact configuration. (a) Alpha-particle loss traces at \(s_0=0.01\) and \(0.25\); the legend gives the complete configuration--line-style combinations. (b) Neoclassical effective ripple for the compact configuration, with W7-X High Mirror retained as the common reference. Unavailable radial samples are omitted.}
\label{fig:conf-compact}
\end{figure}

No alpha-particle loss is observed among the 2000 particles launched from either \(s_0=0.01\) or \(0.25\) during the 0.2~s traces in Fig.~\ref{fig:conf-compact}. The effective ripple is close to W7-X High Mirror. The compact configuration therefore combines a small plasma volume with excellent fast-ion confinement and competitive neoclassical transport.

\begin{table}[!htbp]
\centering
\caption{Geometric and confinement metrics for the compact configuration.}
\label{tab:compact-metrics}
\small
\begin{tabularx}{\linewidth}{@{}l*{5}{>{\centering\arraybackslash}X}@{}}
\toprule
Configuration & \(A\) & \(E\) & Edge well & \(\alpha\) loss, \(s_0=0.01\) & \(\alpha\) loss, \(s_0=0.25\) \\
\midrule
Compact & 4.30 & 5.00 & \(-1.6\times10^{-1}\) & \(\sim0\%\) & \(\sim0\%\) \\
\bottomrule
\end{tabularx}
\end{table}

Compactness means greater fusion-power density and more fusion power per unit magnet and structural volume, reducing the scale of the magnets, vacuum vessel, and other major components required for a given power target. These reductions offer a direct route toward lower material demand and improved plant economics. The same compactness, however, leaves less radial and geometric space for neutron shielding, breeding blankets, divertor structures, heating and diagnostic access, and coil clearances. It also intensifies the coupling between plasma shaping, MHD stability, and coil complexity. The compact result thus defines a high-performance limit rather than a complete reactor solution. These trade-offs motivate the balanced configuration considered next, in which some compactness and vacuum fast-ion performance are exchanged for finite-pressure stability and a credible coil realization.
\FloatBarrier

\subsection{Balanced Configuration}\label{sec:balanced-configuration}

The balanced configuration addresses the trade-offs exposed by the compact case. Its larger aspect ratio, \(A=8.65\), and elongation, \(E=5.78\), provide additional geometric margin while retaining a coherent single-well field structure and improved action invariance. In return, its vacuum fast-ion confinement is weaker than that of the compact configuration. The relevant question is therefore not whether the balanced case reproduces the compact limit, but whether it retains favorable orbit properties while adding finite-pressure stability and coil accessibility.

\begin{figure}[!htbp]
\centering
\includegraphics[width=0.98\linewidth]{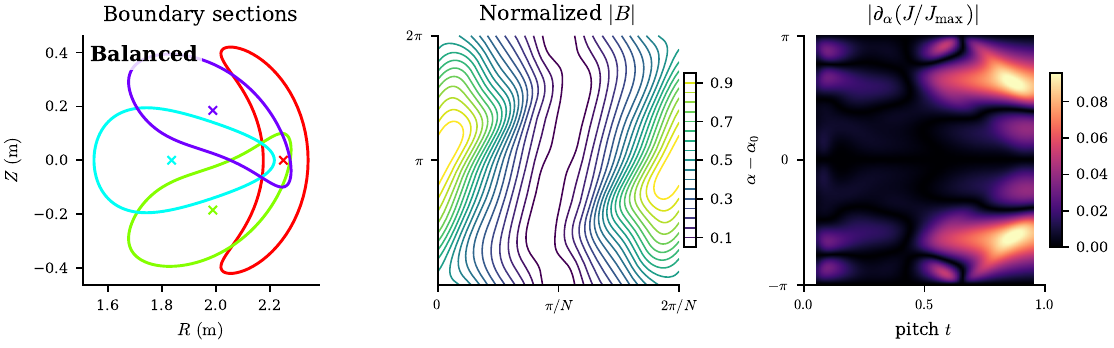}
\caption{Balanced vacuum configuration at \(\rho=1\): boundary sections, normalized Boozer-surface field strength, and \(\lvert\partial_\alpha(J/J_{\max})\rvert\) for the smooth action. The derivative color scale starts at zero and is adapted to this configuration.}
\label{fig:config-balanced}
\end{figure}

\begin{figure}[!htbp]
\centering
\includegraphics[width=0.98\linewidth]{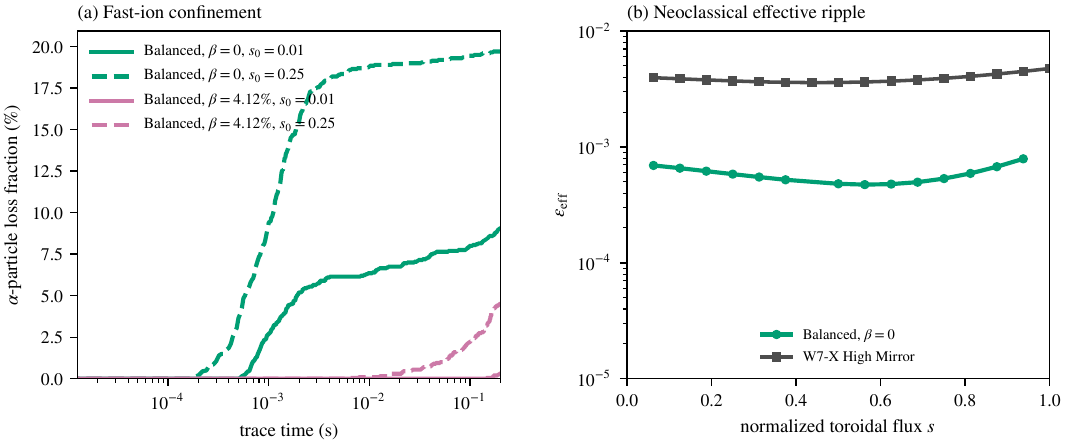}
\caption{Confinement diagnostics for the balanced configuration. (a) Alpha-particle loss traces for the vacuum and \(\langle\beta\rangle=4.12\%\) equilibria; the legend gives every equilibrium--launch-surface color--line-style combination. (b) Neoclassical effective ripple for the balanced vacuum field, with W7-X High Mirror retained as the common reference. Unavailable radial samples are omitted.}
\label{fig:conf-balanced}
\end{figure}

The balanced vacuum configuration loses 9.05\% and 19.75\% of the alpha particles launched at \(s_0=0.01\) and \(0.25\), respectively. At \(\langle\beta\rangle=4.12\%\), the corresponding losses fall sharply to 0.30\% and 4.50\%. This dramatic reduction is directly attributable to the maximum-\(J\) property induced by finite plasma pressure~\cite{helander2014theory}: the diamagnetic self-consistent modification deepens the magnetic well in the plasma core, establishing a negative radial gradient of the second adiabatic invariant (\(\partial J/\partial\rho < 0\)) for trapped particles that strongly stabilizes bounce orbits and suppresses net radial drift. Finite pressure therefore improves fast-ion confinement substantially in this configuration rather than merely preserving the vacuum result. Figure~\ref{fig:conf-balanced}(b) further shows that the balanced vacuum field has substantially lower effective ripple than W7-X High Mirror throughout the available radial interval. The compact and balanced configurations consequently rank differently under fast-ion loss and neoclassical ripple: the compact case gives the lowest observed alpha loss in vacuum, whereas the balanced case gives the lower effective ripple. This distinction shows that neither diagnostic alone captures the complete trapped-orbit response~\cite{escoto2025evaluation}.

\begin{table}[!htbp]
\centering
\caption{Geometric and confinement metrics for the balanced configuration.}
\label{tab:balanced-metrics}
\small
\begin{tabularx}{\linewidth}{@{}l*{5}{>{\centering\arraybackslash}X}@{}}
\toprule
Configuration & \(A\) & \(E\) & Edge well & \(\alpha\) loss, \(s_0=0.01\) & \(\alpha\) loss, \(s_0=0.25\) \\
\midrule
Balanced, \(\beta=0\) & 8.65 & 5.78 & \(+6.5\times10^{-3}\) & 9.05\% & 19.75\% \\
Balanced, \(\beta=4.12\%\) & 8.65 & 5.78 & Not evaluated & 0.30\% & 4.50\% \\
\bottomrule
\end{tabularx}
\end{table}

\begin{figure}[!htbp]
\centering
\includegraphics[width=0.72\linewidth]{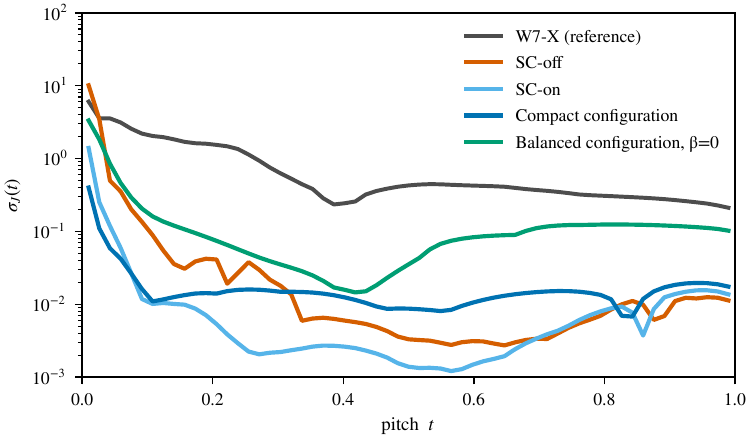}
\caption{Independent hard-cutoff action verification. Normalized \(\sigma_J(t)\) for the controlled SC-off/SC-on pair, compact and balanced configurations, and W7-X, shown on a logarithmic scale. Lower values indicate better omnigenity.}
\label{fig:hard-action}
\end{figure}

A central result is that the favorable action invariance persists when evaluated with the true hard cutoff. The comparison in Fig.~\ref{fig:hard-action} uses
\begin{equation}
\sigma_J(t) = \frac{\max_\alpha J(\alpha,t) - \min_\alpha J(\alpha,t)}{\langle J \rangle_\alpha}
\end{equation}
for all final equilibria. SC-on lies below SC-off over most of the sampled pitch range, and both the compact and balanced configurations remain below W7-X over most pitches. The balanced configuration therefore preserves much of the orbit-action quality demonstrated by the compact case despite the additional stability and engineering demands.

\begin{figure}[H]
\centering
\includegraphics[width=0.72\linewidth]{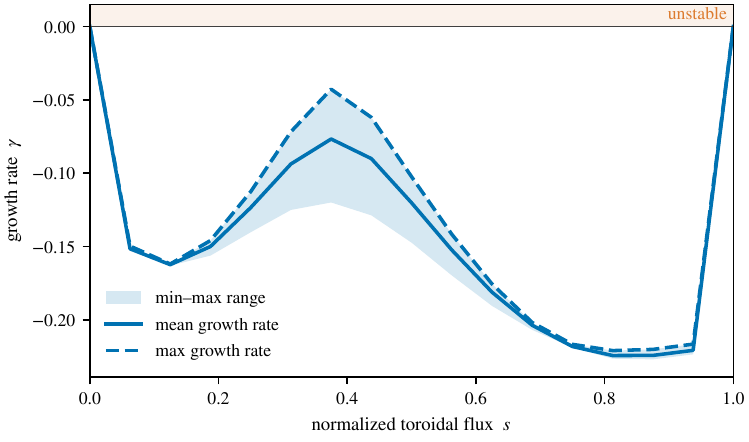}
\caption{Ideal-ballooning stability of the balanced configuration at \(\langle\beta\rangle=4.12\%\). The solid and dashed lines show the mean and maximum growth rates over the sampled field lines, while the shaded band spans the minimum-to-maximum range. The entire profile remains below the marginal boundary \(\gamma=0\).}
\label{fig:balanced-cobra}
\end{figure}

At \(\langle\beta\rangle=4.12\%\), the ideal-ballooning growth-rate profiles remain negative across the entire plasma volume~\cite{sanchez2000cobra} (Fig.~\ref{fig:balanced-cobra}). The least stable region occurs near \(s\approx0.4\), where the maximum growth rate approaches \(-0.043\), but it remains below the marginal boundary. The balanced configuration therefore combines its improved neoclassical transport and finite-pressure fast-ion confinement with a positive ideal-ballooning stability margin.

\begin{figure}[H]
\centering
\includegraphics[width=0.98\linewidth]{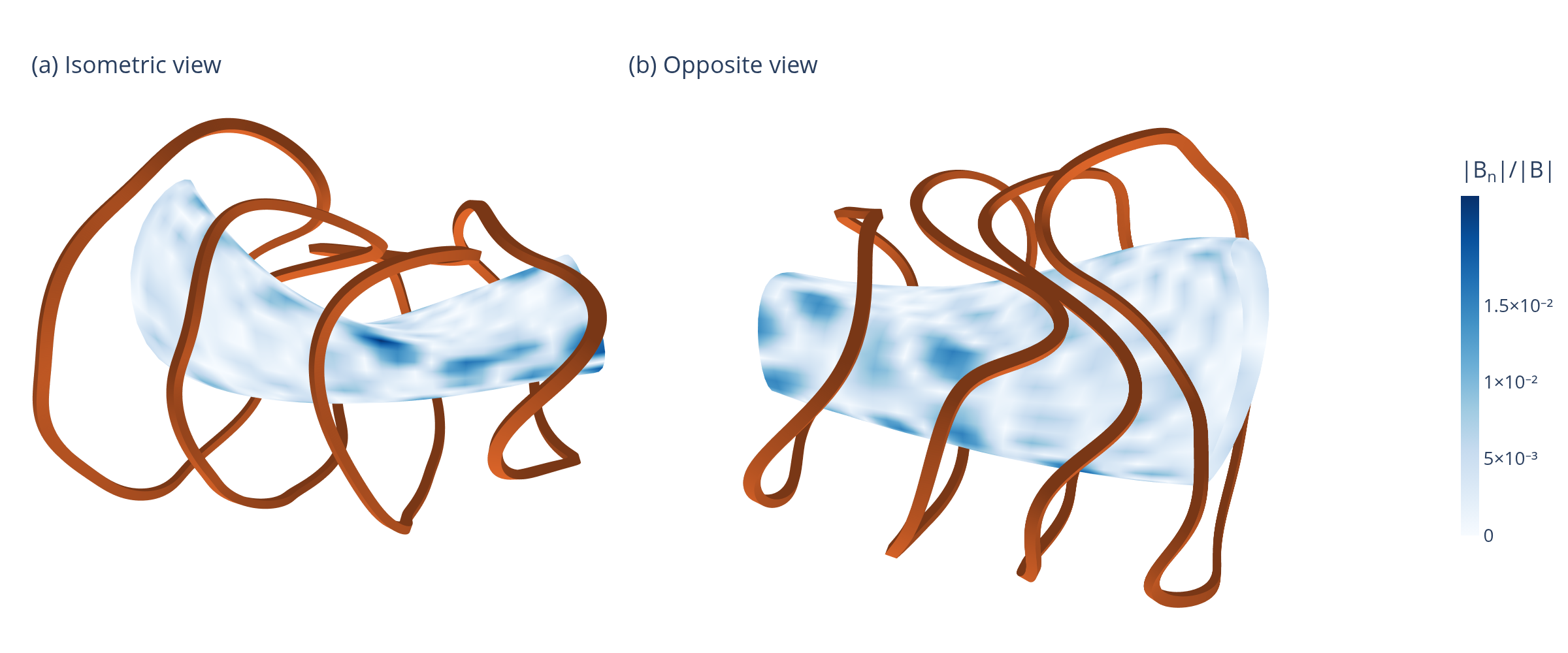}
\caption{Discrete-coil realization of the balanced configuration from two opposing isometric views. Rust-colored rectangular tubes approximate the finite-build conductor cross-sections, and the plasma surface is colored by the local normal-field error \(|B_n|/|B|\).}
\label{fig:balanced-coils}
\end{figure}

The balanced boundary also admits the discrete modular coil set shown in Fig.~\ref{fig:balanced-coils}, comprising four independent coil groups and 32 physical coils after symmetry expansion. The maximum normalized curvature is \(\kappa a=2.15\), the 95th-percentile normalized torsion is \(p_{95}\tau a=2.04\), the minimum normalized coil--coil and coil--plasma distances are 0.458 and 0.752, respectively, and the mean normalized coil length is \(\bar{L}/a=18.6\); these geometric length measures are normalized by the plasma minor radius \(a\). The LCFS area-weighted mean normal-field error is \(<|B_n|/|B|>=4.62\times10^{-3}\). Thus, the configuration retains sub-percent field error and substantial coil--plasma clearance; coil curvature remains its principal engineering limitation.

Taken together, the balanced configuration does not maximize any single performance measure. Instead, it retains hard-action variation below the W7-X reference over most pitches, reduces effective ripple, gains a finite-pressure ideal-ballooning stability margin, and remains compatible with a discrete coil set of sub-percent normal-field error. Its larger aspect ratio reduces the power-density and economic advantages of the compact case, but the added space and stability margin make the configuration more credible as a reactor-facing compromise. The comparison therefore exposes a continuous design trade-off between compactness, confinement, stability, and engineering accessibility rather than a single universally optimal geometry.

\hypertarget{discussion}{%
\section{Discussion}\label{discussion}}

The proposed formulation makes a direct orbit-physics condition usable in gradient-based stellarator optimization. Previously, the defining condition \(\partial J/\partial\alpha=0\) was difficult to optimize because moving bounce boundaries, non-smooth cutoffs, and changing branch connectivity made its gradients unreliable. The present fixed-domain action proxy and analytical derivative turn that condition into a differentiable DESC objective, while the independent hard-action calculation in Fig.~\ref{fig:hard-action} shows that the optimized proxy has the intended physical effect.

The method also shows that a physically meaningful objective must control its own numerical failure modes. Normalized pitch sampling removes field-strength scaling ambiguity. Smooth extrema and softplus cutoffs preserve differentiability. Dynamic field-line sampling maintains consistency with the evolving equilibrium. Soft-connectivity constrains the branch topology required for a global single-branch action residual.

The fully differentiable formulation also brings a practical computational advantage: automatic differentiation provides exact action gradients with respect to all equilibrium degrees of freedom in a single backward pass, avoiding the curse of dimensionality that makes finite-difference gradient evaluation prohibitive for the high-dimensional parameter spaces of stellarator optimization. Although each gradient evaluation is more expensive than a single function call, the overall optimization converges in far fewer iterations than gradient-free alternatives, making direct orbit-constrained design computationally tractable.

Because the action residual directly penalizes \(\partial J/\partial\alpha\) without reference to a prescribed helicity or well shape, the formulation is agnostic to the type of omnigenity being targeted. The same objective can be used to optimize quasi-isodynamic, quasi-helical, or other omnigenous configurations; the magnetic geometry and rotational transform determine which class emerges naturally. The two configurations presented here are four-field-period designs that converge to QI-like topologies, but the method transfers directly to other helicities and field-period numbers without algorithmic modification.

The present residual is applicable to the regime in which the sampled pitch produces one connected trapped interval along a field line; it is not branch resolved. More general piecewise-omnigenous fields can allow particles to switch wells at field-period boundaries, potentially enlarging the design space while retaining favorable transport \cite{velasco2024piecewise,velasco2025exploration,liu2026combining}. These developments motivate a differentiable branch-resolved action objective as the next step: it would enforce \(J\)-invariance branch by branch rather than globally. Smooth identification and tracking of branches through merger and splitting events remain open methodological problems.

In future work, this direct action-based objective can be integrated into single-stage stellarator optimization frameworks that simultaneously optimize coil shapes and plasma equilibrium, enabling end-to-end design of reactor-relevant configurations with direct orbit constraints.

\(J\)-invariance serves as a direct trapped-orbit constraint that complements, rather than replaces, higher-fidelity assessments such as particle tracking, finite-orbit-width studies, turbulence and transport calculations, MHD stability checks, and coil engineering. The value of the present formulation is precisely that it can be combined with those objectives in the same AD workflow; the balanced configuration and its discrete-coil realization provide a first demonstration of this integration. The method thus supplies a new direct trapped-orbit constraint that enriches the existing multi-objective reactor design toolkit.

\hypertarget{conclusions}{%
\section{Conclusions}\label{conclusions}}

We have developed a differentiable action-residual formulation that makes the defining condition of stellarator omnigenity, field-line invariance of the second adiabatic invariant \(J\), usable as a direct objective in gradient-based optimization. Unlike existing approaches that rely on the Cary--Shasharina construction through parameterized targets or symmetry-detecting coordinate mappings, our method evaluates the action residual directly from the equilibrium magnetic field, removing the need for a priori specification of helicity or well topology. By resolving long-standing numerical pathologies associated with moving bounce boundaries, non-smooth cutoffs, and trapped-branch topology, this approach constrains trapped-particle orbits directly from their fundamental physics principle.

Systematic validation using independent hard-cutoff action calculations confirms that optimization of the smooth action proxy yields genuine improvements in physical \(J\)-invariance, rather than purely numerical artifacts. The soft-connectivity regularizer is shown to be essential for this physical interpretability: by enforcing a single-well trapped topology, it eliminates spurious cancellation of variations across disconnected branches, ensuring the residual corresponds to genuine omnigenization of individual bounce orbits.

Applied to four-field-period stellarator design, the method yields two compelling configurations that span the practical design trade space. The compact configuration achieves an aspect ratio of 4.3, approaching super-compact reactor targets, with near-zero alpha-particle losses and \(J\)-invariance exceeding that of Wendelstein 7-X across most pitch values. The balanced configuration trades some of this extreme orbit performance for finite-\(\beta\) ideal-ballooning stability and a realizable modular coil set with sub-percent normal-field error, demonstrating that the direct action objective can be integrated with stability and engineering constraints in a unified optimization workflow. The present formulation assumes a single connected trapped interval per field line, which limits its applicability to globally omnigenous configurations; extension to omnigenous fields with multiple distinct trapping wells remains an open challenge.

Looking forward, extension to branch-resolved action objectives, integration into single-stage coil--plasma co-optimization, and combination with higher-fidelity transport and stability models are natural next steps. By enabling direct gradient-based enforcement of a core adiabatic invariant constraint, this work provides a new tool for the design of next-step stellarator reactors with both high confinement performance and practical engineering viability.

\section*{Acknowledgements}
This work was supported by the National Magnetic Confinement Fusion Energy Program of China (Grant No.~2025YFE03110000), the Anhui Provincial Key Research and Development Project (Grant No.~2023a05020008), the National Natural Science Foundation of China (Grant No.~12405274), the Postdoctoral Fellowship Program of CPSF (Grant No.~GZC20232716), and the Strategic Priority Research Program of the Chinese Academy of Sciences (Grant No.~XDB0790302). The authors declare no conflicts of interest.

\section*{Data Availability}
The data that support the findings of this study are available from the corresponding author upon reasonable request.

\bibliographystyle{unsrtnat}
\bibliography{references}

\end{document}